\documentclass[dvips]{article}
\usepackage{lscape,epsfig}
\usepackage{amssymb}
\usepackage{amsmath}
\usepackage{rotating}
\DeclareSymbolFont{ppa}{OT1}{ppl}{m}{it}
\DeclareMathSymbol{\vv}{\mathalpha}{ppa}{'166}

\begin{document}   

\newcommand{\dd}{\,{\rm d}}
\newcommand{\ie}{{\it i.e.},\,}
\newcommand{\etal}{{\it et al.\ }}
\newcommand{\eg}{{\it e.g.},\,}
\newcommand{\cf}{{\it cf.\ }}
\newcommand{\vs}{{\it vs.\ }}
\newcommand{\zdot}{\makebox[0pt][l]{.}}   
\newcommand{\up}[1]{\ifmmode^{\rm #1}\else$^{\rm #1}$\fi}
\newcommand{\dn}[1]{\ifmmode_{\rm #1}\else$_{\rm #1}$\fi}
\newcommand{\upd}{\up{d}}
\newcommand{\uph}{\up{h}}
\newcommand{\upm}{\up{m}}
\newcommand{\ups}{\up{s}}
\newcommand{\arcd}{\ifmmode^{\circ}\else$^{\circ}$\fi}   
\newcommand{\arcm}{\ifmmode{'}\else$'$\fi}
\newcommand{\arcs}{\ifmmode{''}\else$''$\fi}
\newcommand{\MS}{{\rm M}\ifmmode_{\odot}\else$_{\odot}$\fi}
\newcommand{\RS}{{\rm R}\ifmmode_{\odot}\else$_{\odot}$\fi}
\newcommand{\LS}{{\rm L}\ifmmode_{\odot}\else$_{\odot}$\fi}

\newcommand{\Abstract}[2]{{\footnotesize\begin{center}ABSTRACT\end{center}
\vspace{1mm}\par#1\par
\noindent
{~}{\it #2}}}

\newcommand{\TabCap}[2]{\begin{center}\parbox[t]{#1}{\begin{center}
  \small {\spaceskip 2pt plus 1pt minus 1pt T a b l e}
  \refstepcounter{table}\thetable \\[2mm]
  \footnotesize #2 \end{center}}\end{center}}

\newcommand{\TableSep}[2]{\begin{table}[p]\vspace{#1} 
\TabCap{#2}\end{table}}

\newcommand{\FigCap}[1]{\footnotesize\par\noindent Fig.\  %
  \refstepcounter{figure}\thefigure. #1\par}

\newcommand{\TableFont}{\footnotesize}
\newcommand{\TableFontIt}{\ttit}
\newcommand{\SetTableFont}[1]{\renewcommand{\TableFont}{#1}}

\newcommand{\MakeTable}[4]{\begin{table}[htb]\TabCap{#2}{#3}
  \begin{center} \TableFont \begin{tabular}{#1} #4
  \end{tabular}\end{center}\end{table}}
  
\newcommand{\MakeTableSep}[4]{\begin{table}[p]\TabCap{#2}{#3}
  \begin{center} \TableFont \begin{tabular}{#1} #4
  \end{tabular}\end{center}\end{table}}

\newenvironment{references}%
{
\footnotesize \frenchspacing
\renewcommand{\thesection}{}
\renewcommand{\in}{{\rm in }}
\renewcommand{\AA}{Astron.\ Astrophys.}
\newcommand{\AAS}{Astron.~Astrophys.~Suppl.~Ser.}
\newcommand{\ApJ}{Astrophys.\ J.}
\newcommand{\ApJS}{Astrophys.\ J.~Suppl.~Ser.}
\newcommand{\ApJL}{Astrophys.\ J.~Letters}
\newcommand{\AJ}{Astron.\ J.}
\newcommand{\IBVS}{IBVS}
\newcommand{\PASP}{P.A.S.P.}
\newcommand{\Acta}{Acta Astron.}
\newcommand{\MNRAS}{MNRAS}
\renewcommand{\and}{{\rm and }}
\section{{\rm REFERENCES}}
\sloppy \hyphenpenalty10000
\begin{list}{}{\leftmargin1cm\listparindent-1cm
\itemindent\listparindent\parsep0pt\itemsep0pt}}%
{\end{list}\vspace{2mm}}

\def\TYLDA{~}
\newlength{\DW}
\settowidth{\DW}{0}
\newcommand{\dw}{\hspace{\DW}}
  
\newcommand{\refitem}[5]{\item[]{#1} #2%
\def\REFARG{#3}\ifx\REFARG\TYLDA\else, {\it#3}\fi
\def\REFARG{#4}\ifx\REFARG\TYLDA\else, {\bf#4}\fi 
\def\REFARG{#5}\ifx\REFARG\TYLDA\else, {#5}\fi.}

\newcommand{\Section}[1]{\section{#1}}
\newcommand{\Subsection}[1]{\subsection{#1}}
\newcommand{\Acknow}[1]{\par\vspace{5mm}{\bf Acknowledgements.} #1}
\pagestyle{myheadings}

\newfont{\bb}{ptmbi8t at 12pt}
\newcommand{\xrule}{\rule{0pt}{2.5ex}}
\newcommand{\xxrule}{\rule[-1.8ex]{0pt}{4.5ex}}
\def\thefootnote{\fnsymbol{footnote}}
\begin{center}

{\Large\bf
The Cluster AgeS Experiment (CASE). Variable Stars in the
Globular Cluster M55.
}

\vskip1cm
{
\large J. Kaluzny$^{1}$, I.~B. Thompson$^{2}$, 
W. Krzeminski$^{3}$, K. Zloczewski$^{1}$}
\vskip3mm
{
       $1$ Nicolaus Copernicus Astronomical Center,\\
        ul. Bartycka 18, 00-716 Warsaw, Poland\\
        e-mail: jka@camk.edu.pl, kzlocz@camk.edu.pl\\

      $^{2}$Carnegie Institution of Washington, 813 Santa Barbara Street,
      Pasadena, CA 91101, USA \\
      e-mail: ian@obs.carnegiescience.edu\\

      $^{3}$Las Campanas Observatory,\\ 
      Casilla 601, La Serena, Chile \\
      e-mail: wojtek@lcoeps1.lco.cl\\

}
\end{center}

\Abstract{
We report time-series photometry for 55 variable stars located
in the central part of the globular cluster M55. 
The sample includes 28 newly identified objects of which
13 are eclipsing binaries. Three of these are detached systems 
located in the turn-off region on the cluster color-magnitude diagram.
Two of them are proper motion (PM) members of M55 and are excellent 
candidates for a detailed follow-up study
aimed at a  determination of the cluster age and distance.
Other detached binaries are located along the unevolved part of
the cluster main sequence. 
Most of the variables are cluster blue straggler stars.
This group includes 35 SX~Phe stars, two contact
binaries, and one semi-detached binary. V60 
is a low mass, short period algol with the less massive and cooler
component filling its Roche lobe. The more massive component is
an SX~Phe variable. The  orbital period of V60
increases at a rate of $dP/P$=3.0E-9.  
In addition to numerous variable blue stargglers we also report
the detection of two red stragglers showing periodic variability.  
Both of these are PM members of M55. We note and discuss the observed  
paucity of 
contact binaries among unevolved main sequence stars in M55 and NGC~6752. 
This apparent paucity supports an evolution model in which the formation 
of contact binaries is triggered by stellar evolution at the  
main-sequence turn off.   

}

{\bf Key words:} {\it
stars: dwarf novae -- globular
clusters: individual: M55 --
binaries: eclipsing -- blue stragglers}


\section{Introduction} \label{sect:intro}

M55=NGC~6809 is a nearby globular cluster whose low reddening and
relatively high Galactic latitude 
($(m-M)_{V}=13.87$, $E(B-V)=0.08$, $b=-23.3$~deg; Harris 1996)
makes it an attractive target for detailed studies.  
The photometric survey presented here was conducted 
as a part of the CASE project (Kaluzny et al. 2005) conducted with telescopes
at the Las Campanas Observatory. 
Some results of an earlier photometric study of M55  made 
with the 1.0-m Swope telescope were presented by Pych et al. (2001) 
and  Olech et al. (1999). In these papers we reported the identification 
of several 
SX~Phe stars and the first ever detection of non-radial 
oscillations in RR~Lyr variables. An unreported  result from that survey was 
the detection of a few detached eclipsing binaries.  
We have used the 2.5-m du Pont telescope for a follow-up 
photometic study of these binaries. In addition, we used the du Pont observations 
to search for variables in the inner-most region of the cluster. The central region
of the cluster 
was poorly resolved  on images collected with the Swope telescope. 
In this paper we present light curves and a preliminary analysis
of variables detected on images collected 
over several observing seasons with the du Pont telescope. 

\section{Observations and data analysis}\label{sect:observations}

All images were taken with the 2.5-m
du Pont telescope.  A field of $8.84\times 8.84$ arcmin was 
observed with the TEK5 CCD camera at a scale of 0.259
arcsec/pixel.  Two pointings were used. On most  nights 
the cluster core was positioned 1.5  arcmin south 
and 1.4 arcmin east of the detector center. Below we  call this 
pointing field A. On some nights the cluster core
was positioned at the center of the detector. We  call 
this pointing field B. Note that there was a substantial overlap between 
the two fields and so most  stars were monitored 
on all observing nights. 

Observations were made on a total of 66 nights between 31 May 1997 and 
28 June 2009. Images were taken in two bands with average exposure times of 70~s 
and 120~s for  the $V$ and $B$ filters, respectively. 
The readout time of the detector was 68~s. 
In total, 3388 useful frames in $V$ and
712 frames in $B$ were collected. 
The median seeing was  1.09  and 1.15 arcsec  for 
the $V$ and $B$ bands, respectively.
The stellar photometry was extracted with a modified
version of the image subtraction  package ISIS V2.1 
(Allard \& Lupton 1998; Allard 2000).
{\sc Daophot}, Allstar and Daogrow codes (Stetson 1987, 1990) were used to 
extract lists of point sources and to derive 
aperture corrections for the reference images. For each of the two pointings 
the observed field was divided into a
$4\times 4$ mosaic to reduce effects 
caused by  a spatially variable point spread function. 
Light curves were extracted for  
47138  point like sources detected on the $V$ reference image. 
The instrumental photometry was 
tied to the standard $BV$ system using linear transformations 
based on 328 local standard stars from Stetson (2000).

Fig. 1 shows the rms of individual 
measurements versus average magnitude for $V$-band light curves.
Stars with $V<14.0$ are overexposed on template images.
The photometry has an accuracy of
about 3 mmag  for the brightest unsaturated stars,
decreasing to  14 mmag at $V=18.0$. 

The light curves were checked for variability with
AoV and AOVTRANS algorithms running in the TATRY 
program (Schwarzenberg-Czerny 1996, Schwarzenberg-Czerny \& Beaulieu 2006). 

A total of 54 variable stars located below the horizontal branch on the cluster's
color-magnitude diagram (CMD) were identified in 
our data\footnote{Several brighter RR~Lyr stars were also 
detected but the $V$-band photometry was affected by saturation 
in some frames}. Their equatorial coordinates are listed in Table 1. 
Astrometric solutions were obtained for fields A and B 
using the positions of 662 and 761 stars from the UCAC3 catalog 
(Zacharias et al. 2010), respectively.
Variables V16-V43 were  reported by Pych et al. (2001)
and Kaluzny et al. (2005). 
The remaining 27 objects are new identifications \footnote{
Light curves of the variables discussed in this paper are available
from the CASE archive at http://case.camk.edu.pl}.
Finding charts for these new variables are presented in Fig. 2.   


\begin{table}
\begin{center}
\caption{Equatorial coordinates for M55 variables
}
\begin{tabular}{|l|c|c|l|c|c|}
\hline
   ID &  RA(J2000)& Dec(J2000) & ID & RA(J2000)& Dec(J2000)  \\
      &  [deg]    & [deg]      &    &   [deg]  & [deg]       \\
\hline
16 & 295.03833 & -30.94527 & 45 & 295.03677 & -30.95457 \\
17 & 295.04724 & -30.99056 & 46 & 295.02548 & -30.97280 \\
18 & 295.02864 & -30.94251 & 47 & 295.01940 & -30.96607 \\
19 & 294.99030 & -30.95061 & 48 & 295.01224 & -30.95817 \\
20 & 294.97897 & -30.97283 & 49 & 295.05049 & -30.97099 \\
21 & 294.99278 & -30.98528 & 50 & 295.05603 & -30.99595 \\
22 & 295.03250 & -31.00376 & 51 & 295.04699 & -30.91533 \\
23 & 294.96592 & -30.93159 & 52 & 295.02693 & -30.99111 \\
24 & 294.93953 & -30.93433 & 53 & 295.02178 & -30.98953 \\
25 & 294.96479 & -30.93950 & 54 & 294.90595 & -30.95111 \\
26 & 294.94607 & -30.95970 & 55 & 295.03304 & -30.94736 \\
27 & 294.97521 & -30.96900 & 56 & 295.00456 & -30.95584 \\
29 & 294.92741 & -30.93312 & 57 & 294.99827 & -30.94833 \\
31 & 295.00412 & -30.96597 & 58 & 294.99567 & -30.95132 \\
32 & 294.99224 & -30.97601 & 59 & 295.01676 & -30.97048 \\
33 & 294.97731 & -30.99967 & 60 & 294.99046 & -30.96288 \\
34 & 295.00425 & -31.01080 & 61 & 295.00281 & -30.99124 \\
35 & 294.95985 & -30.92036 & 62 & 294.95751 & -30.90109 \\
36 & 294.95234 & -30.94610 & 63 & 294.95539 & -30.94218 \\
37 & 294.95777 & -30.96207 & 64 & 294.94666 & -30.95341 \\
38 & 294.99525 & -30.97104 & 65 & 294.96326 & -30.98365 \\
39 & 295.04998 & -31.03483 & 66 & 294.94217 & -30.99322 \\
40 & 295.00790 & -30.92753 & 67 & 294.93898 & -31.00914 \\
41 & 295.01230 & -30.97480 & 68 & 295.00426 & -31.01171 \\
42 & 294.99423 & -30.95691 & 69 & 294.97403 & -31.01317 \\
43 & 295.03599 & -30.98135 & 70 & 294.88948 & -30.90194 \\
44 & 295.04360 & -30.96861 & 71 & 295.01376 & -30.98486 \\   
\hline
\end{tabular}
\end{center}
\end{table}

\section{SX Phe stars}\label{sect:sx} 
The sample of variables includes 35 SX~Phe stars. 
Table 2 provides some basic photometric parameters of these stars.
Average magnitudes are given in columns 3 and 5 while
light curves amplitudes are given in columns 4 and 6.  
For multi-mode pulsators the listed periods correspond to the 
largest amplitude mode. As can be seen in Fig. 3,  SX~Phe variables 
constitute most  of the cluster blue stragglers
brighter than $V\approx 17.5$. Few constant stars are located 
in this part of the cluster CMD. This 
point will be discussed further in a forthcoming proper-motion 
study  of the cluster (Zloczewski et al., in preparation). 
The luminosities of SX Phe stars belonging to M55 span a range 
of $\Delta V\approx 1.5$  implying a substantial range of 
stellar mass.
The pulsational characteristics of these stars are worth detailed
modeling given that they share the same metallicity
and distance; this in turn puts strong constraints on models.

The sample includes 8 single mode pulsators. All but one of these 
showed coherent oscillations with constant amplitude over
12 years of observations. The exception is V26 which showed
changes of the period as well as the amplitude.
Light curves of V26 for seasons 1997-2001, 2003-2006 and 2007-2009 can 
be phased with periods of 0.082011545(4), 0.082008684(24) and
0.082011912(13) days, respectively. For the $V$-band  the amplitude
of the variability ranged from 0.098 in the 2004 season to 0.041
in the 2008 season. This is shown in Fig. 4. 
We checked that the observed changes of the period and amplitude
are not due to oscillations in two very closely separated modes.
An explanation of the unusual behavior of V26 is left to experts 
in the field of stellar pulsations. We note that also some of multi-periodic
variables also showed noticeable instabilities of their periods and 
amplitudes over the 12 year interval.

\begin{table}
\begin{center}
\caption{Periods, average magnitudes and amplitudes of SX~Phe stars
in M55}
\begin{tabular}{|l|l|c|c|c|c|c|}
\hline
   ID & P[d]       & $<V>$    & AV    & $<B>$  & AB & Mode$^{a}$  \\
\hline
16 & 0.05341898(1) & 16.974 & 0.006 & 17.331 & 0.008 & s \\
17 & 0.04126184(1) & 17.191 & 0.018 & 17.523 & 0.023 & s \\
18 & 0.04655539(1) & 16.999 & 0.011 & 17.317 & 0.013 & s \\
19 & 0.03823603(2) & 17.326 & 0.012 & 17.635 & 0.016 & s \\
20 & 0.03321202(5) & 17.018 & 0.033 & 17.360 & 0.047 & m \\
21 & 0.13559137(1) & 15.775 & 0.010 & 16.200 & 0.013 & m$^{b}$ \\
22 & 0.04563898(1) & 16.806 & 0.123 & 17.102 & 0.151 & s \\
23 & 0.04140053(1) & 17.231 & 0.018 & 17.551 & 0.022 & s \\
24 & 0.04181954(4) & 17.062 & 0.002 & 17.380 & 0.003 & m \\
25 & 0.09853144(1) & 15.882 & 0.285 & 16.210 & 0.364 & s \\
26 & 0.0820104(2)  & 16.110 & 0.047 & 16.495 & 0.063 & s \\
27 & 0.0410321(1)  & 17.131 & 0.010 & 17.471 & 0.012 & m \\
29 & 0.0343117(2)  & 20.681 & 0.090 & 20.842 & 0.104 & m \\
31 & 0.03884781(5) & 17.278 & 0.016 & 17.628 & 0.025 & m \\
32 & 0.0414872(1)  & 16.953 & 0.037 & 17.275 & 0.046 & m \\
33 & 0.0573482(6)  & 16.400 & 0.055 & 16.725 & 0.041 & m \\
34 & 0.03701723(6) & 17.238 & 0.011 & 17.542 & 0.015 & m \\
35 & 0.0486828(2)  & 16.586 & 0.015 & 16.894 & 0.022 & m \\
36 & 0.03939643(2) & 16.741 & 0.024 & 17.036 & 0.031 & m \\
37 & 0.04379779(6) & 16.949 & 0.017 & 17.239 & 0.023 & m \\
38 & 0.03817392(7) & 16.709 & 0.018 & 17.108 & 0.027 & m \\
39 & 0.0358108(1)  & 17.209 & 0.012 & 17.495 & 0.013 & m \\
40 & 0.03697678(3) & 17.217 & 0.010 & 17.534 & 0.013 & m \\
41 & 0.09033409(1) & 16.536 & 0.040 & 16.826 & 0.049 & m \\
42 & 0.0366654(2)  & 17.180 & 0.018 & 17.423 & 0.025 & m \\
45 & 0.03083606(5) & 16.330 & 0.004 & 16.572 & 0.004 & m \\
47 & 0.0249482(3)  & 16.786 & 0.003 & 17.053 & 0.003 & m \\
48 & 0.0330509(3)  & 16.113 & 0.024 & 16.378 & 0.025 & m \\
52 & 0.0276799(3)  & 16.153 & 0.003 & 16.359 & 0.003 & m \\
55 & 0.0323921(1)  & 17.228 & 0.002 & 17.615 & 0.003 & m \\
57 & 0.02185877(8) & 16.644 & 0.003 & 16.894 & 0.003 & m \\
59 & 0.02624428(7) & 16.650 & 0.007 & 16.897 & 0.012 & m \\
61 & 0.03490324(9) & 17.305 & 0.008 & 17.620 & 0.012 & m \\
63 & 0.05308522(9) & 16.399 & 0.002 & 16.707 & 0.003 & m \\
69 & 0.0450622(2)  & 17.027 & 0.007 & 17.350 & 0.011 & m \\
\hline
\end{tabular}
\end{center}

\vspace*{-0.3cm}
{\footnotesize Note: $^{a}$~single (s) or  multi mode (m) pulsator , 
$^{b}$~period for seasons 2001-2008, for seasons 1999-2001 
$P=0.13552290(2)$.
}

\end{table}

\section{Eclipsing binaries and unclassified variables}

In Table 3 we list the basic properties of variables other than the SX~Phe 
stars discussed above. Ephemerides are given in columns 2 and 3, followed
by $V$ and $B$ magnitudes at maximum light, the amplitude of the variability
for the $V$ filter, the average $B-V$ color and the assigned variability type.
The last column lists the membership status based on proper 
motion measurements (Zloczewski et al., in preparation). 
Figure 5 shows the location of the variables on the cluster CMD.

The sample includes eight eclipsing binaries with periods ranging
from 0.54 to 17.2 days. Phased light curves of these stars are 
shown in Fig. 6.
All but one of these are detached systems. The exception is the semi-detached blue
straggler V60, further discussion of this system is given in section 4.1. 
The detached binaries V44, V54 and V58
are located near the turnoff region of the cluster. V44 and V54  are  PM 
members of M55 while PM status of V58 remains undetermined. 
These systems deserve a detailed 
study aimed at a determination of the absolute parameters of their components.
This in turn can provide information about the age and distance of 
M55 (Paczy\'nski 1997). Binaries V62 and V66 are located slightly above the 
lower main sequence of the cluster. The former is a PM member 
of M55 while there is no PM information for V66. According to its PM 
V49 is  a field object. Its location on the cluster 
CMD indicates that it is a foreground  binary. Its light curve shows large
season-to-season changes which is not unusual given the late spectral type
inferred from the observed color. We do not have PM information for 
V68 but its location on the CMD indicates that it is a field object
located behind the cluster. Its color and magnitude are appropriate
for an early F-type binary belonging to the Sagittarius dwarf galaxy.
We note that Pych et al. (2001) detected  3 SX~Phe stars in the M55 field
which belong to the Sagittarius galaxy. These variables are located in 
the same part of the cluster CMD as V68 and are also expected to
be stars of spectral type F.

Five objects were classified as contact binaries. Their phased light
curves are shown in Fig. 7.

No PM information is available for V56 and V70.
Examination of their light 
curves leads to the conclusion  that the individual components 
of V56 and V70 are located on the cluster  main sequence.
Therefore, both systems are likely members of M55.
The large amplitude of the light curve of V70 indicates
that it has a mass ratio close to unity. 

The blue straggler V53 is a PM 
member of M55 while there is no PM information for the blue straggler
V46. We note that light curves of both binaries show asymmetric maxima.
These asymmetries were observed in all 9 observing seasons and so it is unlikely that  
they are caused by stellar spots. The asymmetries are more likely 
related to some phenomena resulting from mass transfer 
in V46 and V53. 

The variable V51 has a small but measurable proper motion: 
$\mu_{\alpha}=0.512\pm 0.295$, $\mu_{\delta}=4.441\pm 0.163$.
It is definitely a field object not related to the cluster.  
Its light curve resembles light curves of low amplitude
contact binaries. However, on the color-period diagram (Rucinski 2002) 
the variable is located outside of the region occupied by contact binaries.
With $(B-V)_{0}=0.15$ and $P=0.45$~d the star is too blue
to be an ordinary contact binary\footnote{Throughout this paper 
we use $E(B-V)=0.115$ following Dotter et al. (2010)}. 
Given its location on the CMD and the shape of the light curve 
V51 is possibly a very distant, close but non-contact  binary with  
low orbital inclination. Alternatively it might be a
binary sdB system showing orbital modulation of its light.

The variables V64 and V65 lie slightly to the 
red of the subgiant branch on the cluster CMD. Both stars are 
PM members of M55. As can be seen in Fig. 8 they show
systematic season-to-season changes of their average luminosities. 
At the same time their light curves show coherent periodic variability 
with changing amplitudes, as demonstrated in Fig. 9. 
This coherence of the observed periodic variations suggests that  
V64 and V65 are binary systems. Bassa et al. (2008) 
identified V65 as the likely optical conterpart to the X-ray source CX7.
They also note that it is a candidate active magnetic binary.
Time domain and phased light curves of V67 are presented in Figs. 8 \& 9. 
The object is located right on the 
main sequence of M55 and it is PM member of the cluster.
The amplitude of the periodic light variations is  small. The star
is probably either a contact binary observed at low inclination or an ellipsoidal
variable. In the later case the short orbital period suggests that
one of the components is a degenerate star. 
Variable V71 is a likely optical 
counterpart to the X-ray source CX8 (Bassa et al. 2008). The star
is located at the base of the subgiant branch on the cluster CMD
and it is PM member of M55.
As  is shown in Fig. 8  the average luminosity of V71 changes on 
a yearly time scale. Periodic
modulation of the light curve with $P\approx8.04$~d was observed in 
some seasons, as is demonstrated in Fig. 9. 
Spectroscopic observations can help to answer  questions about the 
nature of the four objects discussed briefly in this paragraph. 

As can be seen in Fig. 8 the
cataclysmic variable V43=CV1 showed 2 outbursts during our observations.
In quiescence the variable is relatively red and occupies a position
close to the  main-sequence on the cluster CMD. No periodic variability could be 
detected in quiescence.

The variable   V50 is a field star based on PM measurements.
Its light curve does not show any periodicity.
However, as can be seen in Fig. 8, the average luminosity of V50 showed 
noticeable changes on a time scale of 10 years. The star is a likely
optical counterpart to the X-ray source CX29 (Bassa et al. 2008)

\begin{table}
\begin{center}
\caption{Basic data for eclipsing binaries and unclassified  variables
in M55
}
\begin{tabular}{|l|l|l|c|c|c|c|c|c|}
\hline
   ID &  P[d] & T$_{0}$    & $V_{\rm MAX}$ & $B_{\rm MAX}$ & $\Delta V$ & $<B-V>$ & Type & M \\
      &       & HJD-2450000 &               &               &            &         &    &    \\
\hline
43 &   -            &    -       &  19.1 &  19.45 &  -   &  0.68  & CV$^{a}$& Y \\
44 &   2.166123(5)  & 2076.8618  & 17.843& 18.360 &  0.25&  0.538&  EA$^{b}$& Y\\
46 &   0.31941843(1)&  599.3135  & 16.462& 16.725 &  0.35&  0.260&  EW$^{c}$& - \\
49 &   1.95902(4)   &  599.5426  & 19.15 & 20.24  &  0.47&  1.06 &  EA      & N  \\ 
50 &   -            &    -       & 17.08 & 17.98  &  0.30&  0.90 &   ?      & N\\     
51 &   0.4522456(4) &  599.0813  & 18.445 &18.71  &  0.07 & 0.269&  EW      & N\\ 
53 &   0.32524682(1)&  599.2113  & 16.545& 16.796 &  0.05&3 0.251&  EW      & Y\\
54 &   9.26917(1)   & 2869.7370  & 18.300& 18.857 &  0.36&  0.56 &  EA      & Y\\
56 &   0.29186192(1)&  599.1822  & 17.74 & 18.28  &  0.36&  0.51 &  EW      & -\\
58 &   8.6205:      &  2769.8231 & 18.01 & 18.56  &  0.20&  0.55:&  EA      & -\\
60 &   1.1830214(7) &  599.1814  & 16.83 & 17.24  &  1.78&  0.41 &  EA      & Y\\
62 &   1.2358474(5) &  599.2222  & 20.26 & 21.08  &  0.49&  0.82 &  EA      & Y\\ 
64 &   12.945:      & 1339.5:   & 17.02 & 17.77  &  0.18&  0.77 &  ?        & Y\\
65 &   5.587538:   & 3887.7:    & 17.22 & 18.10  &  0.19&  0.88 &  ?        & Y\\
66 &   0.5496634(7) &  599.4780  & 20.22 & 21.12  &  0.41&  0.92 &  EA      & -\\
67 &   0.121062:    & 3861.9:    & 18.31 & 18.84  &  0.12&  0.53 &  ?       & Y\\
68 &   0.53857748(3)&  599.2259  & 20:65:& 20.92: &  0.52:&  0.24: &  EA    & -\\
70 &   0.22287334(4)&  599.1016  & 19.40 & 20.12  &  0.72&  0.76 &  EW      & -\\
71 &   8.0399:      &  601.5:    & 16.87 & 17.57  &  0.69&  0.13 & ?        & Y \\
\hline
\end{tabular}
\end{center}
\vspace*{-0.3cm}
{\footnotesize Note: $^{a}$~cataclysmic variable (Kaluzny et al. 2005), 
$^{b}$~detached or semi-detached binary,
$^c$~contact binary}
\end{table}

\subsection{The Blue straggler V60}

The blue straggler V60 is a PM member of M55.
Figure 10 shows phased $B$ and $V$ light curves  for the
observing seasons 2006-2009. The depth of the 
primary eclipse increases from about $1.78$~mag in the $V$ band to about
$2.17$~mag in the $B$ band indicating a large difference of the  effective
temperatures of the components of V60.  The orbital period of the system is unstable. 
The $O-C$ diagram presented in Fig. 11 includes
moments of primary eclipses based on the du Pont observations as well as some
measurements based on unpublished CASE observations collected with the 
Swope telescope. The relevant data are listed in Table 4. 
The $O-C$ diagram indicates that orbital period of the variable 
increases at the rate of $dP/dt\approx 3.0E-9$. This can be interpreted as 
the  result of conservative  mass transfer in the system and is consistent with 
a semi-detached configuration reported below.
An accurate determination of the 
parameters of the binary is hampered by the lack of spectroscopic information
about the  mass ratio. Nonetheless, we  obtained a  preliminary
solution of the light curves shown in Fig. 10 using the Wilson-Devinney
code (Wilson \& Devinney 1971) working under control of the  Phoebe  
utility (Pr\^sa \& Zwitter 2005). We used portions of the 
data from observing seasons 2006-2009. Light curves were phased separately
for each of these seasons using the values 
of $T_{0}$ listed in Table 4.  
The mass ratio $q=m_{s}/m_{p}$ 
was treated as an adjustable parameter. Several trials were made starting 
with different values of $q$ assuming a detached configuration. 
The solution indicates a semi-detached configuration with 
a mass ratio $q=0.20$, inclination $i=87.1$~deg and $R_{p}/R_{s}=0.81$.
Thus the binary
is a classical algol with the less massive and cooler component filling its 
Roche lobe.  
The luminosity ratio at quadratures is $(L_{p}/L_{s})_{V}=3.46$ 
and $(L_{p}/L_{s})_{B}=5.95$. 
The resulting  apparent (reddened) colors of the components are
$(B-V)_{p}=0.31$  and $(B-V)_{s}=0.90$. 
With $V_{p}=17.11$ the 
primary  of V60 is located  among the SX~Phe  pulsating 
blue stragglers on the M55 CMD  (see Fig. 3). Low amplitude pulsations
with a period of about 0.03~d are in fact easily seen in the light 
curve of the binary. To check if these oscillations are coherent
we used the following procedure. A second order polynomial was fitted to 
the out-of-eclipse sections of the nightly $V$ light curves 
from the 1999 observing season. The residual light curves were then merged
and searched for periodicity with TATRY. A strong signal was detected at 
a period of 0.03087(7). The residual light curve phased with such a 
period is shown in Fig. 12. 

The light curve of V60 is   symmetric
and shows only minor season-to-season changes. 
The system is apparently free from significant  photometric disturbances 
caused by mass transfer or chromospheric activity. Therefore, a reliable and 
accurate determination of its geometrical parameters should be 
possible once the mass ratio is constrained by spectroscopy. 
The binary is worth a more detailed follow-up study
including  a determination of its
absolute parameters. The primary component of V60 is a Pop II  star
with ${\rm [Fe/H]}=-1.9$ and a mass  noticeably exceeding masses
of turnoff stars in globular clusters. As such it may provide an interesting 
benchmark for models of low metallicity stars.

\section{Summary and Discussion}\label{sect:summary}
According to a proper motion study conducted by Zloczewski et al. 
(2010; in preparation) most of the variables reported in this paper 
are members of M55. As we have already noted several of them deserve a spectroscopic 
study.
In the case of eclipsing binaries such a study should lead to a determination 
of their absolute parameters and distances. At this time the absolute parameters 
are known for very few Pop II stars. Six stars of different masses 
constituting three detached binaries from the cluster turnoff should lay 
on the same isochrone. Hence, knowing their absolute parameters
one may conduct an interesting test of stellar models at low metallicity.
Spectroscopic observations of several other variables would also be valuable, 
confirming with high confidence the
membership status of objects that are candidate PM members of the cluster. 
The systemic velocity of M55 is $V_{r}=174.8$~km/s.

We note the apparent paucity of contact binaries
among stars on the cluster main-sequence. Two systems were detected among 
the blue stragglers and one at the turnoff at $V=17.7$. 
Two of these three are PM members of M55 and the PM status for third
one is unknown. In contrast just 
one contact binary was found below the turnoff. This is variable
V70 with $V=19.4$ and an unknown membership status. The sample of surveyed stars 
is strongly dominanted by lower main sequence stars. We have examined the variability status
of 13775 stars with $V<18.0$ and 25530 stars with $18.0<V<19.5$,
where $V\approx18.0$ corresponds to the turnoff level. 
As can be seen in Fig. 1 the accuracy of our differential 
photometry is such that it would be easy to detect any potential 
contact binaries with $V<19.5$ and $\Delta V>0.1$. A similar
lack of contact binaries among main-sequence stars is also
observed in NGC~6752. Our recent survey of this nearby globular cluster
revealed two contact systems among the blue stragglers and none among main-sequence
stars (Kaluzny \& Thompson  2009). Our results indicate that the formation of W UMa stars  is driven 
mainly by nuclear evolution of stars. Once the more massive component of a 
close but initially detached binary reaches the turnoff its expansion
causes Roche lobe overflow and eventually leads to a contact configuration
of the system. After mass reversal the binary may end up as a blue straggler
with a configuration that depends on the efficiency 
of mass and angular momentum loos. The observed lack of contact binaries
among unevolved main sequence stars in GCs supports the evolution model
advocated by Gazeas \& Stepien (2008).


\begin{table}
\begin{center}
\caption{Times of primary minima of V60.
The $O-C$ residuals are calculated with ephemeris from
Table 3.
}
\begin{tabular}{|r|r|c|c|}
\hline
   E &  HJD         & sigma & $O-C$    \\
      & -245 0000   &       &       \\
\hline
 -3415 &   615.72698 &  0.00016 &  0.00405 \\
 -3092 &   997.84749 &  0.00004 &  0.00081 \\
 -2797 &  1346.84636 &  0.00019 & -0.00551 \\
 -2180 &  2076.77070 &  0.00100 & -0.00305 \\
 -1853 &  2463.62201 &  0.00016 & -0.00499 \\
 -1514 &  2864.66824 &  0.00024 & -0.00554 \\
 -1267 &  3156.87576 &  0.00023 & -0.00573 \\
 -1225 &  3206.56427 &  0.00008 & -0.00717 \\
 -1197 &  3239.68670 &  0.00100 & -0.00488 \\
  -655 &  3880.88230 &  0.00020 & -0.00060 \\
  -323 &  4273.64582 &  0.00024 &  0.00038 \\
  -264 &  4343.44210 &  0.00100 &  0.00261 \\
     0 &  4655.75744 &  0.00009 &  0.00603 \\
    37 &  4699.52850 &  0.00100 &  0.00691 \\
   293 &  5002.38290 &  0.00022 &  0.00707 \\
\hline
\end{tabular}
\end{center}
\end{table}


\Acknow{
Research of JK \& KZ  is supported by the grant MISTRZ 
from the Foundation for the Polish Science and by the grant 
N N203 379936 from the Ministry of Science and Higher Education.
I.B.T. acknowledges the support of NSF grant AST-0507325.
We thank Pawel Moskalik for help with time series analysis of  
variable V26.
}

\newpage

\begin{figure}[htb]
\centerline{\includegraphics[width=120mm]{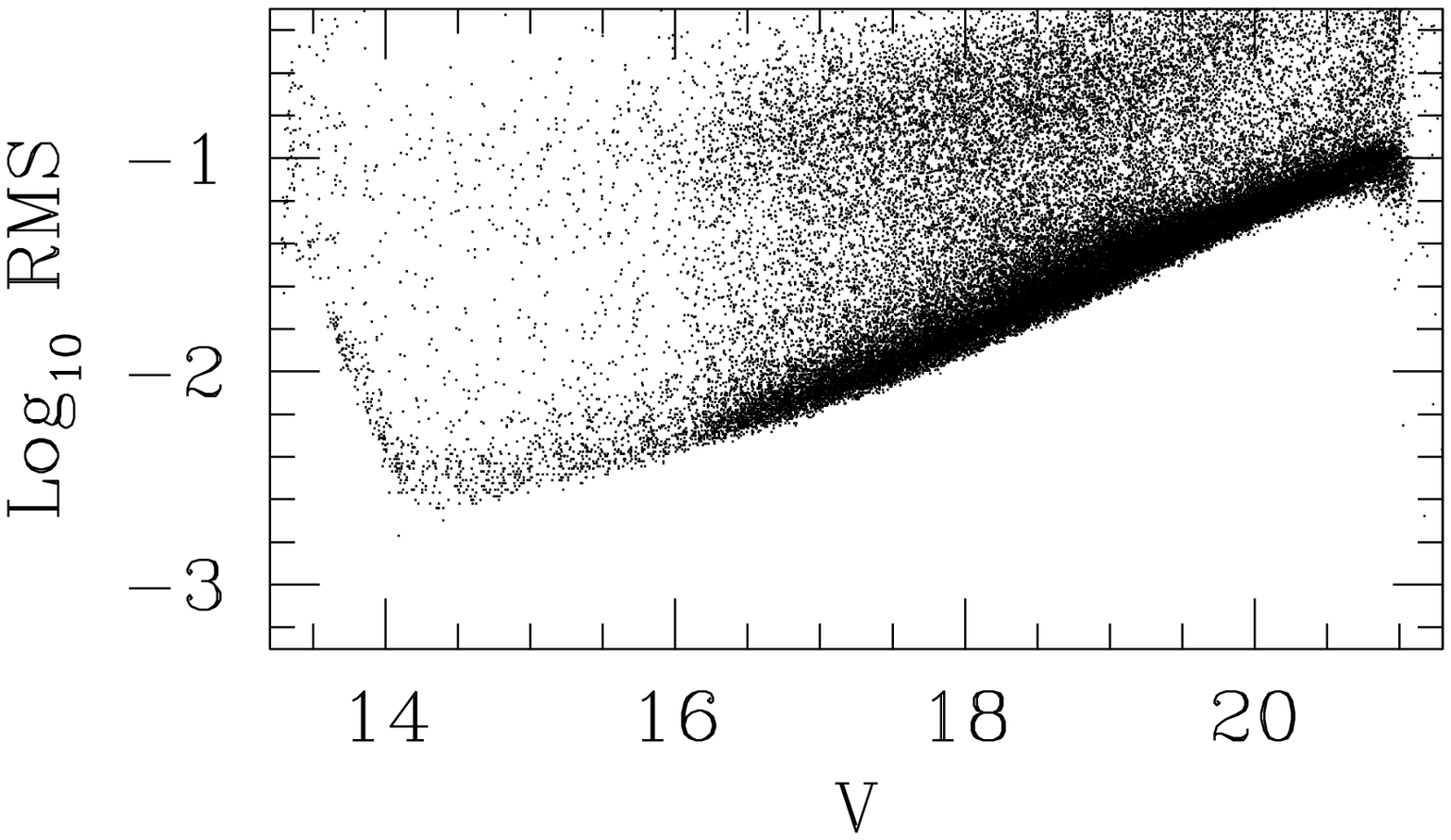}}
\caption{\small Standard deviation vs. average $V$ magnitude for
light curves of stars from M55 field.}
\end{figure}

\begin{figure}[htb]
\centerline{\includegraphics[width=120mm]{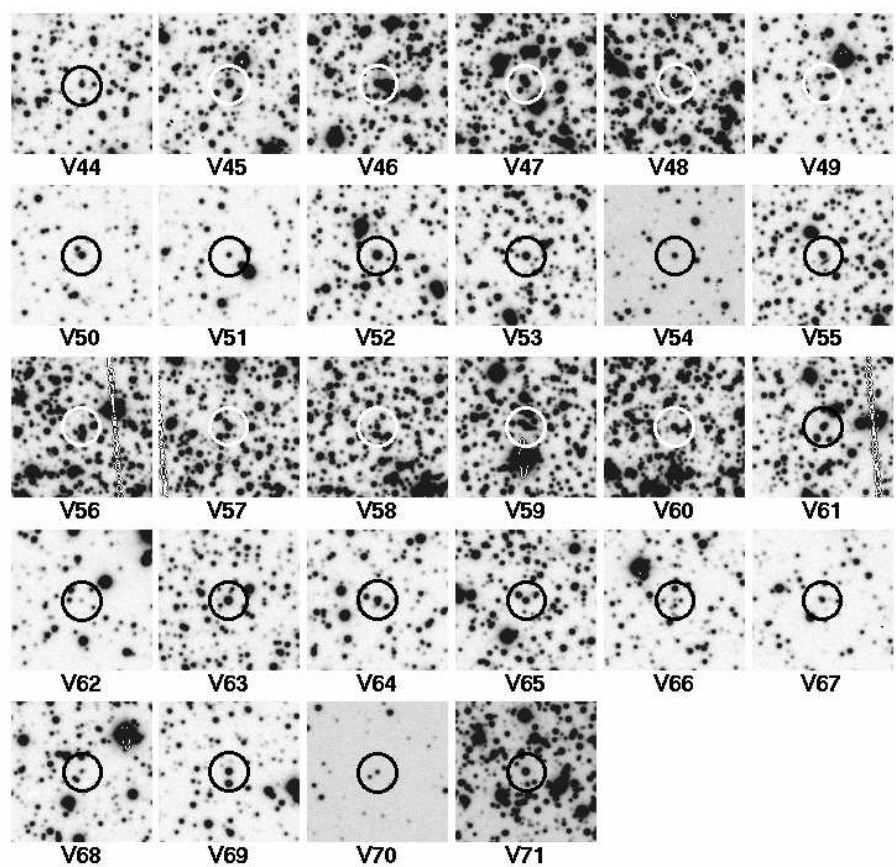}}
\caption{\small Finding charts for variables V44-70.
Each chart is 30 arcsec on a side: north is up and east to the left.}
\end{figure}

\begin{figure}[htb]
\centerline{\includegraphics[width=120mm]{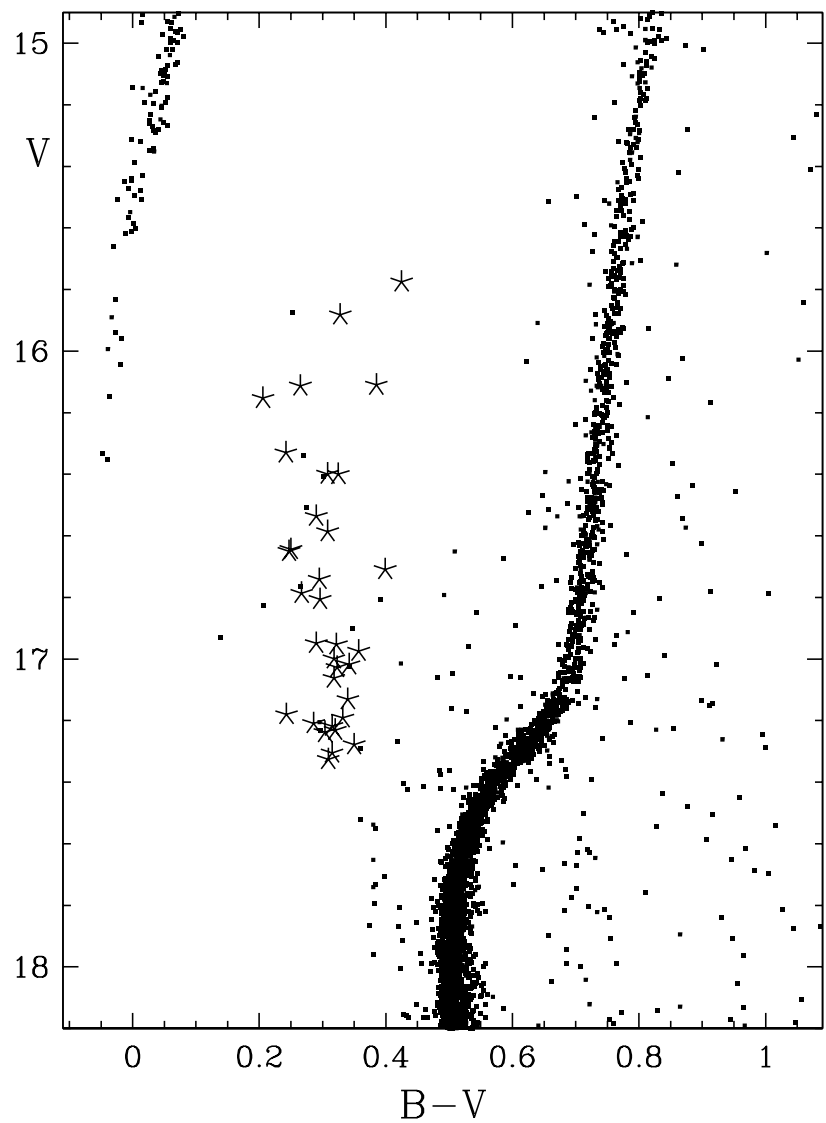}}
\caption{\small Color-magnitude diagram for M55, with positions 
of SX~Phe variables marked with asterisks.}
\end{figure}

\begin{figure}[htb]
\centerline{\includegraphics[width=120mm]{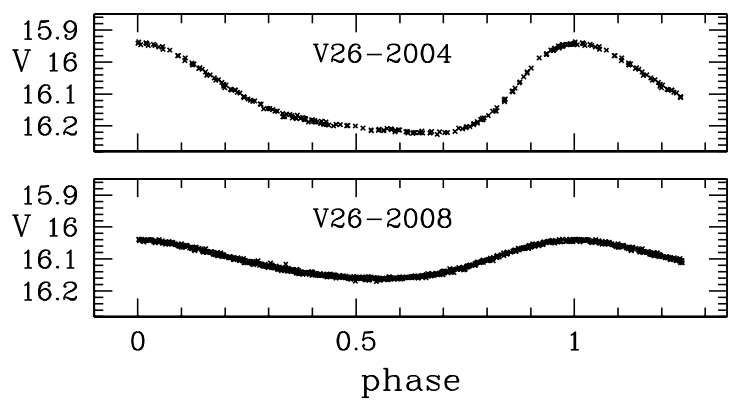}}
\caption{\small Phased light curves of SX~Phe variable V26
for observing seasons 2004 and 2008.
}
\end{figure}

\begin{figure}[htb]
\centerline{\includegraphics[width=120mm]{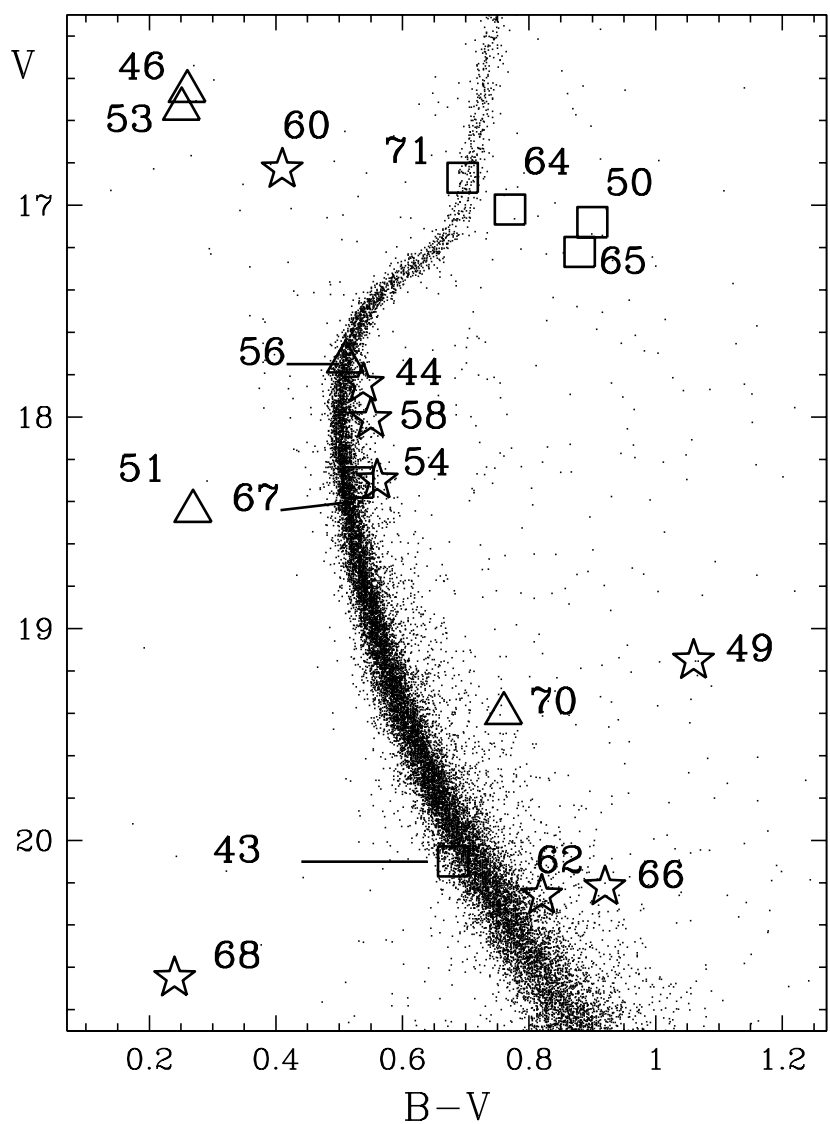}}
\caption{\small Color-magnitude diagram for M55 with positions of 
variables marked. Stars - detached or semi-detached binaries; triangles - 
contact binaries; squares unclassified variables and dwarf nova V43=CV1 . 
}
\end{figure}

\begin{figure}[htb]
\centerline{\includegraphics[width=120mm]{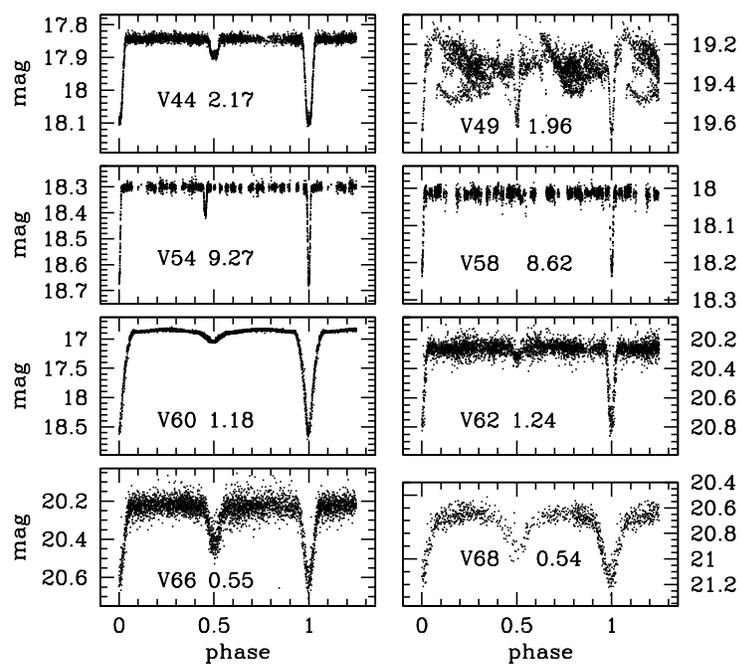}}
\caption{\small Phased $V$ light curves of variables
classified as detached or semi-detached binaries.
Inserted labels give names of variables followed
by their period in days.
}
\end{figure}

\begin{figure}[htb]
\centerline{\includegraphics[width=120mm]{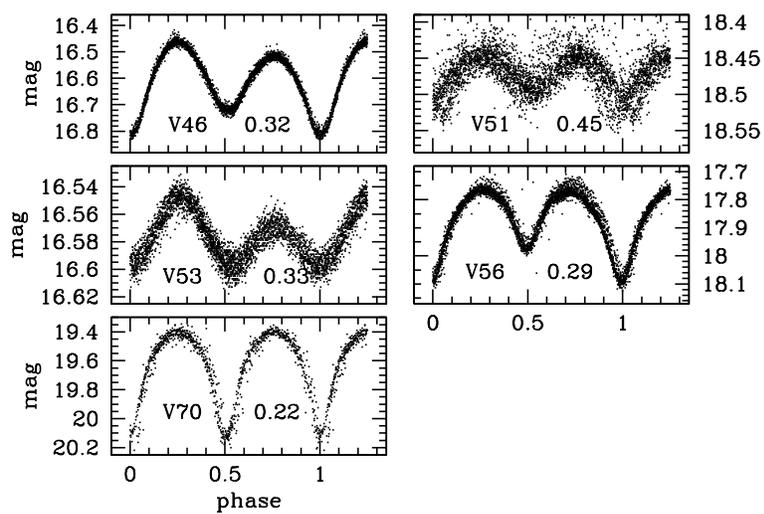}}
\caption{\small Phased $V$ light curves of five variables classified
as contact binaries. Inserted labels give names of variables followed
by their period in days.}
\end{figure}

\begin{figure}[htb]
\centerline{\includegraphics[width=120mm]{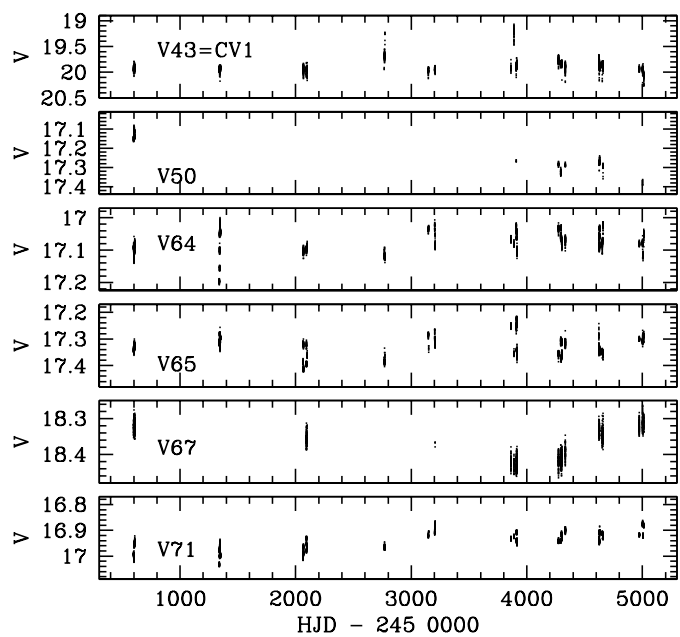}}
\caption{\small Light curves of variables V43, V50, V64, V65 and V66
}
\end{figure}

\begin{figure}[htb]
\centerline{\includegraphics[width=120mm]{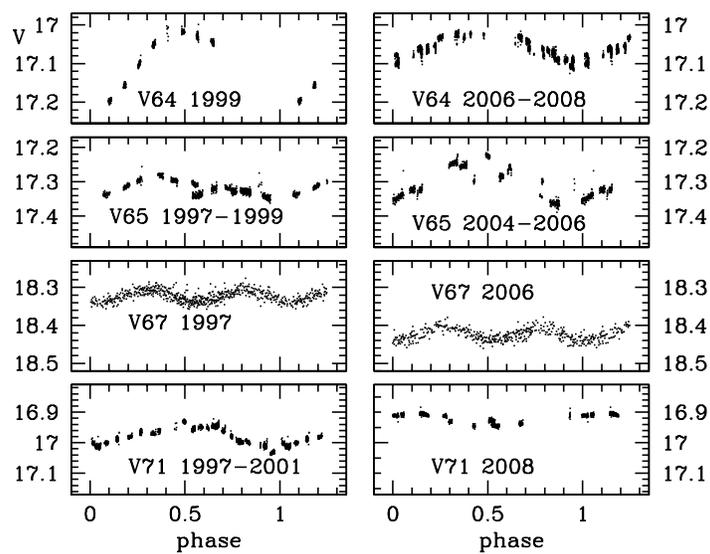}}
\caption{\small Phased $V$ light curves of periodic variables
V64, V65, V67 and V71. Inserted labels give names of variables and 
respective observing season(s).
}
\end{figure}

\begin{figure}[htb]
\centerline{\includegraphics[width=120mm]{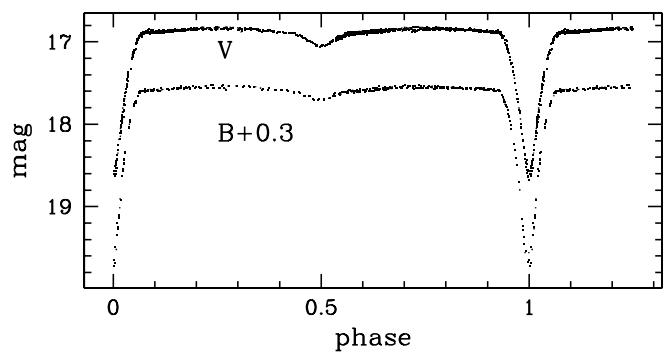}}
\caption{\small Phased  light curves of an eclipsing
blue straggler V60. Note shift applied to $B$  curve.
}
\end{figure}

\begin{figure}[htb]
\centerline{\includegraphics[width=120mm]{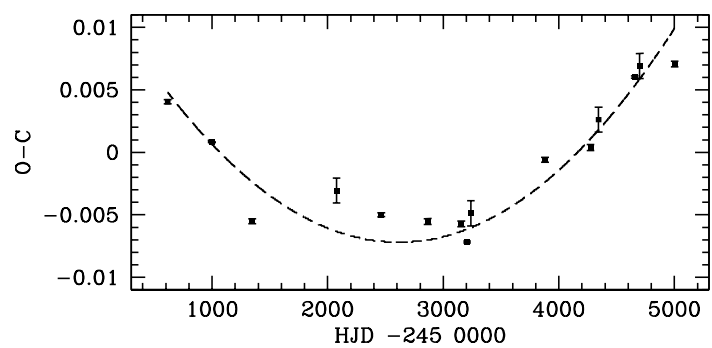}}
\caption{\small 
The $O-C$ diagram of V60 constructed with the linear ephemeris 
listed in Table 3. Dashed line shows formal parabolic fit to the data.
}
\end{figure}

\begin{figure}[htb]
\centerline{\includegraphics[width=120mm]{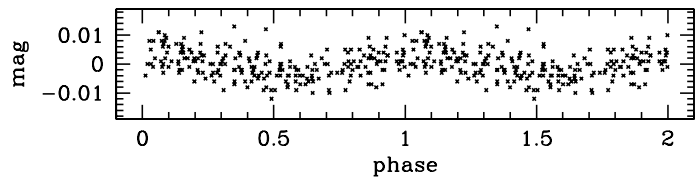}}
\caption{\small 
Residual light curve of V60 phased with a period of 0.03087~d.
See text for details.
}
\end{figure}

\end{document}